\documentclass[preprint,12pt]{article}
\usepackage{graphicx} 
\usepackage{graphics,cite,amssymb,amsmath,epsfig,float,psfrag,enumerate}
\usepackage[usenames]{color}%,dvips]{color}
\usepackage{rotating}

\usepackage{epsfig,amssymb}
\usepackage{latexsym}
\usepackage{verbatim}

\RequirePackage{lineno}

\newcommand{\bq}{\begin{equation}}
\newcommand{\eq}{\end{equation}}
\newcommand{\bqa}{\begin{eqnarray}}
\newcommand{\eqa}{\end{eqnarray}}
\newcommand{\ben}{\begin{enumerate}}
\newcommand{\een}{\end{enumerate}}
\newcommand{\bc}{\begin{center}}
\newcommand{\ec}{\end{center}}
\newcommand{\bqb}{\begin{eqnarray*}}
\newcommand{\eqb}{\end{eqnarray*}}

\newcommand{\qsl}{q\hskip-0.21cm\slash}
\newcommand{\qpsl}{q'\hskip-0.29cm\slash}

\newcommand{\eesl}{\epsilon\hskip-0.21cm\slash}

\newcommand{\beq}{\begin{eqnarray}} 
\newcommand{\eeq}{\end{eqnarray}}

\begin{document}
%%%%%%%%%%%%%%%%%%%%%%%%%%%%%%%%%%%%%%%%%%%%%%%%%%%%%%%%%%%%%%%%%%%%%%%%%%%%%%%

%\setpagewiselinenumbers
%\modulolinenumbers[5]
%\linenumbers

\vspace{1cm}

\vspace*{.5cm}

\begin{center}

{\large\bf  New Physics signals from measurable polarization asymmetries at LHC}

\vspace*{.8cm}

%\mbox{\large M. Beccaria$^1$, A. Djouadi$^{2,3}$, G. Macorini$^4$, G. Panizzo$^5$ and C. Verzegnassi$^5$}
\mbox{\large M. Beccaria$^1$, G. Macorini$^2$, G. Panizzo$^3$ and C. Verzegnassi$^3$}
\vspace*{.8cm}

$^1$
Dipartimento di Matematica e Fisica ``Ennio de Giorgi'', Universit\`a del Salento and INFN, Sezione di Lecce, 
Via Arnesano 1, 73100 Lecce, Italy.\\
%$^2$
%Laboratoire de Physique Th\'eorique, Universit\'e Paris XI et CNRS,
%Orsay, France.\\
%$^3$
%Theory Unit, CERN, 1211 Gen\`eve 23, Switzerland.\\
$^2$ 
Niels Bohr International Academy and Discovery Center, Blegdamsvej 17 DK-2100
Copenhagen, Denmark\\
$^3$
Dipartimento di Fisica , Universit\`a di Trieste and INFN Sezione di 
Trieste, \\ Strada Costiera, 11
I - 34151 Trieste,  Italy. \\

\end{center}

\vspace{1.4cm}

\setcounter{page}{1}

%\tableofcontents

\begin{abstract}

%\section*{Abstract }

We propose a new type of Z polarization asymmetry in bottom-Z
production at LHC that should be realistically measurable and would
provide the determination of the so-called $A_b$ parameter, whose available
measured value still appears to be in disagreement with the Standard
Model prediction. This polarization can be measured independently of a
possible existence of Supersymmetry. If Supersymmetry is found, a second
polarization, i.e. the top longitudinal polarization in top-charged Higgs
production, would neatly identify the $\tan \beta$  parameter. In this case, the
value of $A_b$ should be in agreement with the Standard Model. If
Supersymmetry does not exist, a residual disagreement of $A_b$ from the
Standard Model prediction would be a clean signal of New Physics of
``non Supersymmetric'' origin.

\end{abstract}
%\end{frontmatter} 

% \section{ Introduction} \label{sec:intro}
{\bf  1. Introduction}
\medskip

 The polarized bottom-Z forward-backward asymmetry has been defined several years ago \cite{Blondel:1987gp}, and considered to be  the best way of measuring, in a theoretical SM approach, a combination of the polarized bottom-Z couplings. The definition of this quantity was chosen as
 \begin{equation}\label{eq:polAFBdef}
 A_{FB}^{b,pol} = \frac{(\sigma_{e_L^- b_F}-\sigma_{e_R^- b_F})-(\sigma_{e_L^- b_B}-\sigma_{e_R^- b_B})}{\sigma_{e_L^- b_F}+\sigma_{e_R^- b_F}+\sigma_{e_L^- b_B}+\sigma_{e_R^- b_B}}~,
 \end{equation}
where $b_{F,B}$ indicates forward and backward outgoing bottom quarks respectively (a polarization degree of the incoming beam $= 1$ is for simplicity assumed). At the Z peak one may easily verify that
\begin{equation}
A_{FB}^{b,pol}=\frac{3}{4} \frac{g_{Lb}^2-g_{Rb}^2}{g_{Lb}^2+g_{Rb}^2}~,
\end{equation}
where $g_{L,Rb}$ are the couplings of a left and right handed bottom to the $Z$. Calling
\begin{equation} \label{eq:Ab}
A_b =  \frac{g_{Lb}^2-g_{Rb}^2}{g_{Lb}^2+g_{Rb}^2}~,
\end{equation}
one finds that 
\begin{equation} \label{eq:synthafbpol}
A_{FB}^{b,pol}=\frac{3}{4} A_b.
\end{equation}
The quantity $A_b$ appears also in an unpolarized transition from an electron-positron to a $b-\bar{b}$ pair. One finds in that case that the unpolarized forward-backward $b$ asymmetry at the $Z$ peak can be written as
\begin{equation} \label{eq:AFBunpol}
A_{FB}^{b}=\frac{3}{4} A_e A_b~,
\end{equation}
where $A_e$ is the longitudinal electron polarization asymmetry \cite{Lynn:1986ir}

\begin{equation} \label{eq:Ae}
A_e =  \frac{g_{Le}^2-g_{Re}^2}{g_{Le}^2+g_{Re}^2}
\end{equation}
and equations \eqref{eq:AFBunpol} \eqref{eq:Ae} can be extended to a different final quark antiquark couple $f\bar{f}$, giving
\begin{equation}\label{eq:AFBf}
A_{FB}^{f}=\frac{3}{4} A_e A_f~,
\end{equation}
where $A_f$ is the analogue of $A_b$ eq \eqref{eq:Ab} with $f$ replacing $b$.
The direct measurement of $A_b$, that requires the use of initially longitudinally polarized electrons, was performed at SLAC \cite{Abe:1994tv,Abe:2000dq}, and the result was found to be in good agreement with the Standard Model prediction, that is \cite{Baak:2012kk}
\[A_b^{SM,~th}=0.93464^{+0.00004}_{-0.00007}~.\]
Later, LEP1 performed a number of unpolarized measurements at the Z peak from which the value of $A_b$ was derived.
This was obtained from eq. \eqref{eq:AFBf} and found to be in severe disagreement, at the $3\sigma$ level, with the SM prediction \cite{ALEPH:2005ab}. 
This result was in a certain sense unexpected, because the relative decay rate of the Z into bottom pairs
$R(b)=\Gamma(Z\rightarrow b \bar{b})/\Gamma(Z\rightarrow hadrons)$ provided a value
\begin{equation} \label{eq:Rb}
R_b\simeq g_{Lb}^2+g_{Rb}^2
\end{equation}
in perfect agreement with the SM prediction \cite{ALEPH:2005ab}. Accepting the LEP1 result for $A_b$, a search started of possible new physics models that might have cured the disagreement. In particular, it was concluded that a conventional MSSM was unable to save the situation \cite{ref:MSSMunable}. This conclusion remained problematic, since no extra measurements of $A_b$ were eventually performed, and the emerging picture seems definitely unclear. In addition to the previous statements, a new feature has now appeared. In a very recent important paper \cite{Freitas:2012sy}, a SM calculation of $\sin^2 \theta_W^{eff,b}$ and $R_b$ has been redone including higher order previously neglected effects. 
The result is that the SM theoretical prediction for $A_b$ and $R_b$ are now different \cite{Baak:2012kk}, in the sense that the disagreement of $A_b$ has been slightly ($\sim 2.5\sigma$) reduced, while a new disagreement ($\sim 2.4 \sigma$) for $R_b$ has appeared. 
Certainly, a new measurement of $A_b$ and $R_b$ would therefore represent an undoubtedly relevant 
improvement of our understanding. 
In this paper, we discuss the possibility of a measurement of $A_b$.

In a recent paper \cite{Beccaria:2012xw}, we have defined a certain polarization asymmetry $A_Z^{pol,b}$ to be measured in bottom-Z production at LHC, and shown that this would represent a possibility of measuring the $A_b$ quantity. From a theoretical point of view, this asymmetry exhibits the remarkable properties of being QCD scale and PDF set choice independent, which would represent a quite remarkable feature. From the realistic experimental point of view, this asymmetry should be derived from the experimental determination of the so called \emph{polarization fractions} (see for example \cite{Stirling:2013muo} and references therein) of the Z boson in $bZ$ associated production, known to be affected by intrinsically large \emph{systematic} uncertainties. The aim of this paper is that of proposing an alternative quantity, proportional to $A_b$, measurable in the same process of bottom-Z production at LHC, that would be experimentally clean being eventually limited in precision only by statistical uncertainties. This will be done in the following Section 2 of the paper. In Section 3, the possible relevance of the measurement of another polarization asymmetry, the top longitudinal polarization in top-charged Higgs production, will be discussed in the case of a SUSY discovery. The importance of a measurement of the Z polarization in bottom-Z production with or without Supersymmetry will be finally discussed in Section 4.

%The aim of this paper is that of proposing an alternative quantity, measurable in the same process of %bottom-Z production at LHC, that would be experimentally clean 

%\section{ Helicity amplitudes and $A_{FB}^b$} \label{sec:afb}
\bigskip
{\bf 2. Helicity amplitudes and $A_{FB}^b$} 
\medskip

\begin{figure}
	%\vspace{0.1\textheight}
	\centering
	\includegraphics[trim=4.5cm 11.5cm 4.5cm 12.0cm, clip=true,scale=0.8]{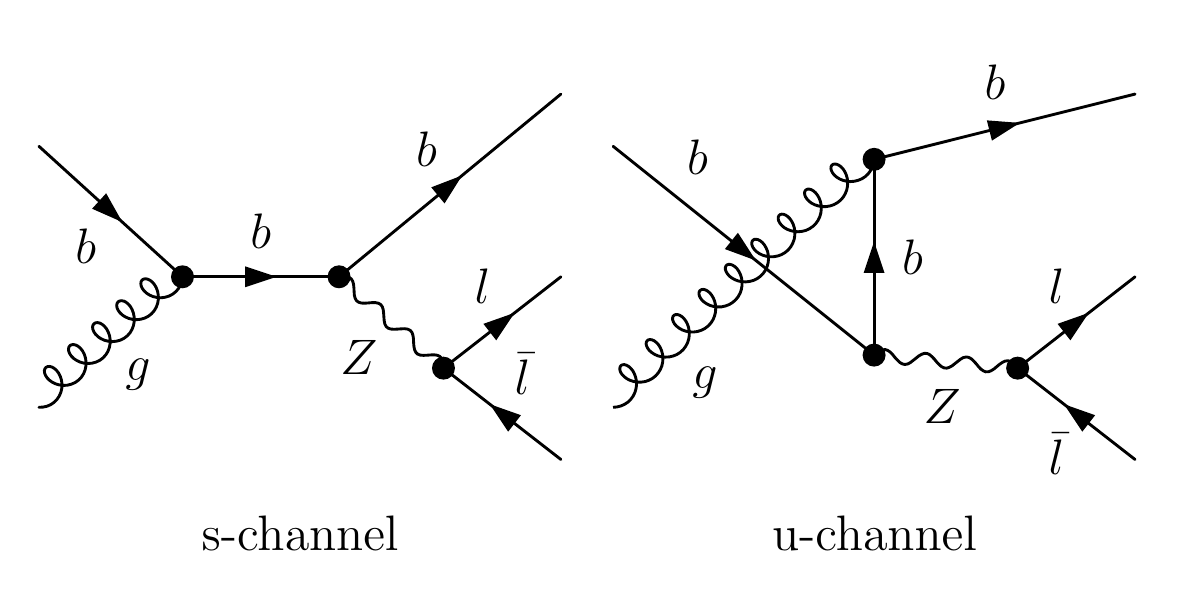}	
				\caption{Leading Order Feynmann diagrams for the process $bg\leftrightarrow bl\bar{l}$}\label{fig:diag}%
\end{figure}

The process of associated production of a single b-quark and a Z boson with its subsequent decay into a lepton-antilepton pair, represented in Figure~\ref{fig:diag},
is defined at parton level by subprocesses $bg\to b l \bar{l}$ involving two Born 
diagrams with bottom quark exchange in the $s$-channel and in the $u$-channel. 
The interaction vertexes involved in the diagrams of Figure~\ref{fig:diag}~
are defined as follows 
\bq
gqq:~~ig_s\eesl \left({\lambda_c^{k}\over2} \right )~~~~~~~~
Zff: -ie \gamma^\mu [g^L_{Zf}P_L+g^R_{Zf}P_R] \equiv -ie \gamma^\mu [g^j_{Z_f}P_j] .
\eq
Therefore, the Born invariant amplitude is given by
\bqa
A^{Born}(bg\to bZ \to bl\bar{l} )&=&4 \pi \alpha g_s\left({\lambda_c^{k}\over2}\right)
\bar u(b')~[\gamma^\mu \{g^j_{Zb}P_j\} {(\qsl+m_b) \over s-m^2_b} ~\eesl 
\nonumber\\
&&
+{\eesl (\qpsl+m_b) \gamma^\mu \over u-m^2_b}
\{g^j_{Zb}P_j\}~]~u(b)~ D_Z(p_Z^2)~ ~\bar{u}(l)~ \gamma_\mu \{g^j_{Zl}P_j\} ~ v(\bar{l}),
\eqa
\noindent
where $\epsilon$, $\lambda_c^{k}$ are the gluon polarization vector and
colour matrix,  $p_l +p_{\bar{l}}  \equiv p_Z $ , $D_Z(p_Z^2) $ is the usual Z effective propagator, $q=p_b+p_g=p_Z+p'_b$, $s=q^2$, $q'=p'_b-p_g=p_b-p_Z$,
$u=q^{'2}$ and  with the kinematic decompositions in the center of mass frame (all fermion massless)

\bq
p_b=(p;0,0,p)~  , ~~~~
p'_b=(p_1;0,p_1\sin\theta_1,p_1\cos\theta_1) \footnote{An additional azimuthal angle for $b'$ would manifest itself only through overall phase factors in the amplitudes.} ,
\eq 
\bq
p_g=(p;0,0,-p)~ , 
\eq
\bq
p_l=(p_2;p_2\sin\theta_2 \sin \phi_2,p_2 \sin\theta_2 \cos \phi_2 ,p_2\cos\theta_2) ~,
\eq
\bq p_{\bar{l}}=(p_3;p_3\sin\theta_3 \sin \phi_3,p_3 \sin\theta_3 \cos \phi_3 ,p_3\cos\theta_3) ~ ,
\eq
\bq
\epsilon (g)=\left (0;{\lambda_{g}\over\sqrt{2}},-~{i\over\sqrt{2}},0 \right )~,
\eq 
where the set of variables ${p_i,~\theta_i,~\phi_i}$ is underconstrained for clarity of notation; a more appropriate set of variables that fullfill $p_b+p_g=p'_b + p_l +p_{\bar l}$ is found rotating the three momenta of the leptons in a new `helicity' frame,
\begin{comment} Figure \ref{fig:coordsys},  
\begin{figure}
	%\vspace{0.1\textheight}
	\centering
	\includegraphics[scale=0.6]{graphs/coordinatesystem_hyde.pdf}	
				\caption{Center of mass coordinate system chosen to compute the helicity amplitudes. In blue the production plane, the black arrow shows the final state bottom quark momentum, in red the decay plane.} \label{fig:coordsys}%
\end{figure} 
\end{comment}
in which the polar axis is the direction of $b'$ and the azimuthal angle is measured from the normal to the production plane (i.e. the one spanned by the colliding and decaying bottom quarks momenta\footnote{The ambiguity coming from the orientation of the normal to the production plane will be canceled after integration over the azimuthal angle in the definition of observable quantities.}). The rotation matrix between the two coordinate systems is
\begin{equation}
R_{\theta_1}=
\begin{pmatrix}
1 & 0 & 0\\
0 & \cos \theta_1 & -\sin \theta_1\\
0 & \sin \theta_1 & \cos \theta_1  
\end{pmatrix}~,
\end{equation}          
from which one can define the polar angles $\theta_{l},\theta_{\bar{l}}$ and the azimuthal angle $\phi'$
\begin{eqnarray*}
p_l^{hf}&=(p_2;p_2\sin\theta_{l} \sin \phi',p_2 \sin\theta_{l} \cos \phi' ,p_2\cos\theta_{l}) ~,\\
p_{\bar{l}}^{hf}&=(p_3;-p_3\sin\theta_{\bar{l}} \sin \phi',-p_3 \sin\theta_{\bar{l}} \cos \phi' ,p_3\cos\theta_{\bar{l}}) ~ .
\end{eqnarray*}
In this frame the coplanarity of the final particles is manifest through the dependence on the same variable $\phi'$ for both leptons. Energy conservation leads, in this frame and for massless particles, to simple formulas for the energies of the final particles (\{$\theta_{l},\theta_{\bar{l}}\}^{h}\equiv\{\theta_{l},\theta_{\bar{l}}\}/2$):
\begin{eqnarray}
p_1 &= p\left(1- \cot (\theta_{\bar{l}}^h) \cot (\theta_{l}^h)\right),\\
p_2 &= p \cos (\theta_{\bar{l}}^h) \csc (\theta_{l}^h)
   \csc (\theta_{\bar{l}}^h+\theta_{l}^h),\\
   p_3&= p \csc (\theta_{\bar{l}}^h) \cos (\theta_{l}^h) \csc (\theta_{\bar{l}}^h+\theta_{l}^h)~,
\end{eqnarray}
which make manifest the (maximal) domain of integration
\[\theta_{l} \in [0,\pi]\text{, } \theta_{\bar{l}} \in [\pi-\theta_{l},\pi]  ~.\]
The introduction of this reference frame is motivated by the cleaner form the matrix elements assume there. 
In the massless case, the helicity amplitudes can be expressed as
\[\mathcal{M}_{\lambda_{b} \lambda_{g};~\lambda_{b'} \lambda_{l} \lambda_{\bar{l}}} 
\equiv \delta_{\lambda_{b} \lambda_{b'}} \delta_{\lambda_{l} \bar{\lambda}_{\bar{l}}} ~ \mathcal{M}_{ \lambda_{g};~\lambda_{b'} \lambda_{l}}~,\]
where $\lambda_f=\pm{1\over2}\equiv\pm$, $\lambda_g=\pm1\equiv \pm$ and $\lambda_i\equiv-\bar{\lambda}_i$.
Modulo a common factor 
\[\mathcal{M}_{ \lambda_{g};~\lambda_{b'} \lambda_{l}} \equiv \left(D_Z(p_Z^2)~16\sqrt{2}~\pi \alpha ~ g_s \lambda_c^{k}\right) F_{ \lambda_{g};~\lambda_{b'} \lambda_{l}}~,\]
the non vanishing helicity amplitudes factors read:
\begin{eqnarray}
F_{+++}=& - i ~(g^R_{Zb}g^R_{Zl})  e^{i \phi'} \sqrt{\frac{p_1 p_2 p_3}{p}} 
\frac{\cos \theta_{\bar{l}}^h \sin \theta_{l}^h}{\cos \theta_1^{h}} ~,\\
F_{++-}=&  i ~(g^R_{Zb}g^L_{Zl})  e^{i \phi'} \sqrt{\frac{p_1 p_2 p_3}{p}} 
\frac{\cos \theta_{l}^h \sin \theta_{\bar{l}}^h}{\cos \theta_1^{h}} ~,\\
 F_{-++}=&  i~ (g^R_{Zb}g^R_{Zl})  e^{-i \phi'}  \sqrt{\frac{ p_1 p_2 p_3}{ p}}
 ~\frac{\sin \theta_{l}^h}{\cos (\theta_{\bar{l}}^h+\theta_{l}^h)}~ 
\frac{\cos \theta_{l}^h }{ \cos \theta_1^{h}} 
\left( \cos \theta_1^{h} \sin \theta_{\bar{l}}^h -e^{i \phi'} \sin \theta_1^{h} \cos \theta_{\bar{l}}^h \right)^2~,\\
F_{-+-}=& - i~ (g^R_{Zb}g^L_{Zl})  e^{-i \phi'} \sqrt{\frac{ p_1 p_2 p_3}{ p}}
 ~\frac{\sin \theta_{\bar{l}}^h }{\cos (\theta_{\bar{l}}^h+\theta_{l}^h)}~
 \frac{\cos \theta_{\bar{l}}^h }{ \cos \theta_1^{h}} 
\left( \cos \theta_1^{h} \sin \theta_{l}^h +e^{i \phi'} \sin \theta_1^{h} \cos \theta_{l}^h \right)^2~,
 \end{eqnarray}
while the other four can be derived by these by parity conjugation, that in our conventions is represented by complex conjugation together with the switch $g^L_{Zf} \leftrightarrow g^R_{Zf}$. As an example  
\[F_{---}=  i ~(g^L_{Zb}g^L_{Zl})  e^{-i \phi'} \sqrt{\frac{p_1 p_2 p_3}{p}} 
\frac{\cos \theta_{\bar{l}}^h \sin \theta_{l}^h}{\cos \theta_1^{h}} ~.\]
Note that formulas related by switch of the lepton helicities are related one to each other by the replacements
\begin{eqnarray}
\left(\theta_{l}\leftrightarrow \theta_{\bar{l}},\phi'\to \phi'+\pi \right)& \equiv \qquad l\leftrightarrow\bar{l}~,\\
g^L_{Zl} ~\leftrightarrow ~ g^R_{Zl}~.&
\end{eqnarray}

 From these formulas one can build the total cross section by introducing the usual flux factor and the convolution with the relevant partons density functions for the proton. For our purposes it suffices to
define the squared amplitude summed over the initial state helicities as
\begin{eqnarray*}
\rho_{\lambda_{b'}\lambda_{l}} &\equiv \sum_{\lambda_{g}}  \vert \mathcal{M}_{ \lambda_{g};~\lambda_{b'} \lambda_{l}} \vert^2
\end{eqnarray*}
and to identify 
\[ \rho_{++} +\rho_{--} \equiv  \left( g_{Lb}^2 g_{Ll}^2 + g_{Rb}^2 g_{Rl}^2\right) f(\theta_l^h,\theta_{\bar{l}}^h,\theta_1,\phi')\]
(one can check that actually in the sum in the RHS  the couplings factorize out). The complete unpolarized squared amplitude can now be simply written as
 \begin{eqnarray}  
 \left | \mathcal{M}\right |^2 &=&  \left( g_{Lb}^2 g_{Ll}^2 + g_{Rb}^2 g_{Rl}^2\right) f(\theta_l^h,\theta_{\bar{l}}^h,\theta_1,\phi')  
 + \left( g_{Lb}^2 g_{Rl}^2 +  g_{Rb}^2 g_{Ll}^2 \right) \bar{f}(\theta_l^h,\theta_{\bar{l}}^h,\theta_1,\phi') \\ &\equiv & c_+ ~\tfrac{f + \bar{f}}{2} ~ + c_- ~\tfrac{f - \bar{f}}{2}~,\label{eq:totunpcs}
\end{eqnarray} 
where $\bar{f} \equiv f|_{l\leftrightarrow\bar{l}}~$. In the last line \eqref{eq:totunpcs}, the two terms have definite symmetry properties under $l\leftrightarrow\bar{l}$, with coefficients
 \begin{eqnarray*}
 c_+ &=& \left(  g_{Lb}^2 + g_{Rb}^2 \right)\left( g_{Ll}^2 + g_{Rl}^2\right) ~,\\
 c_- &=& \left(  g_{Lb}^2 - g_{Rb}^2 \right)\left( g_{Ll}^2 - g_{Rl}^2\right) ~,\\
 &\tfrac{c_-}{c_+}~ = A_b A_l & .
\end{eqnarray*}
This allows us to extract ($c_-$) $c_+$  simply measuring (anti) symmetrized combination of cross sections in  kinematic domains related one to each other under exchange of the two leptons angles. The simplest choice in the CM frame is 
\begin{eqnarray}
\mathcal{D}_\pm &\equiv  \theta_l \gtrless \theta_{\bar{l}}~.
\end{eqnarray}
To be more explicit, note that the condition $\theta_l \gtrless \theta_{\bar{l}}$ translates in the $Z$ rest frame to the experimentally simpler condition of Forward/Backward lepton momentum respect to the bottom momentum versor. This finally leads to the definition of  $A_{FB}^{b,LHC}$ 
\bq
A_{FB}^{b,LHC} \equiv \frac{\sigma(\mathcal{D}_F)-\sigma(\mathcal{D}_B)}{\sigma(\mathcal{D}_F)+\sigma(\mathcal{D}_B)}~,
\eq
where the reference axis is the $b$ momentum in the $Z$ rest frame. From \eqref{eq:totunpcs} this quantity will be proportional, modulo a kinematic factor $k$, to the LEP $A_{FB}^b$
\begin{equation} \label{eq:AFBLHC}
A_{FB}^{b,LHC} = k ~ A_{FB}^{b}~,
\end{equation}
where $FB$, as already emphasized, has different meaning in the two expressions. 

\begin{figure} 
\centering{
\includegraphics[scale=.5]{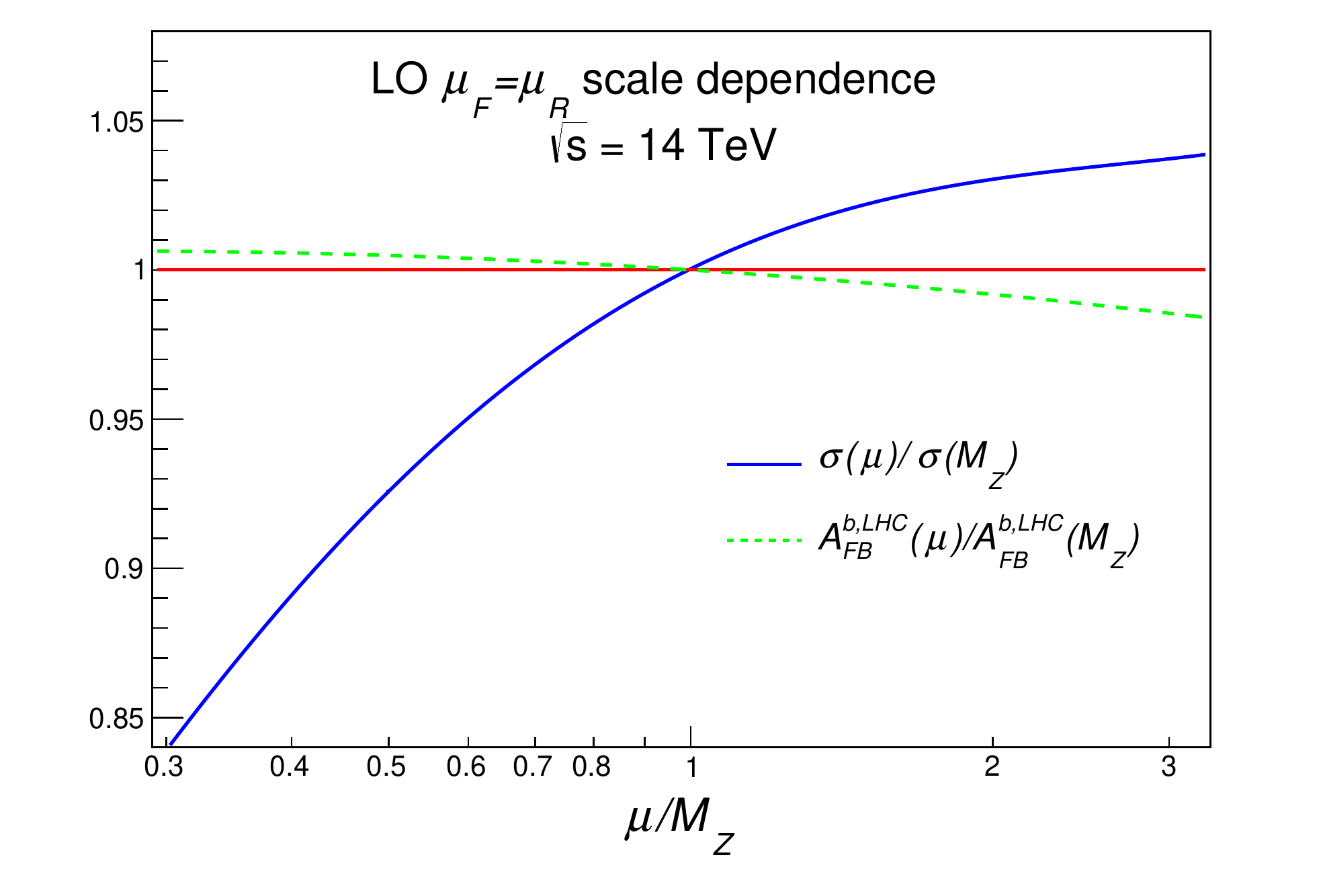}}
\caption{Comparison between the LO $\mu_F=\mu_R=k M_Z$ scale variation dependencies of the total cross section
and our asymmetry. }\label{fig:scale}
\end{figure}

\begin{figure}
\centering
\includegraphics[trim=0.0cm 0.0cm 0.0cm 2.0cm, clip=true,scale=0.6]{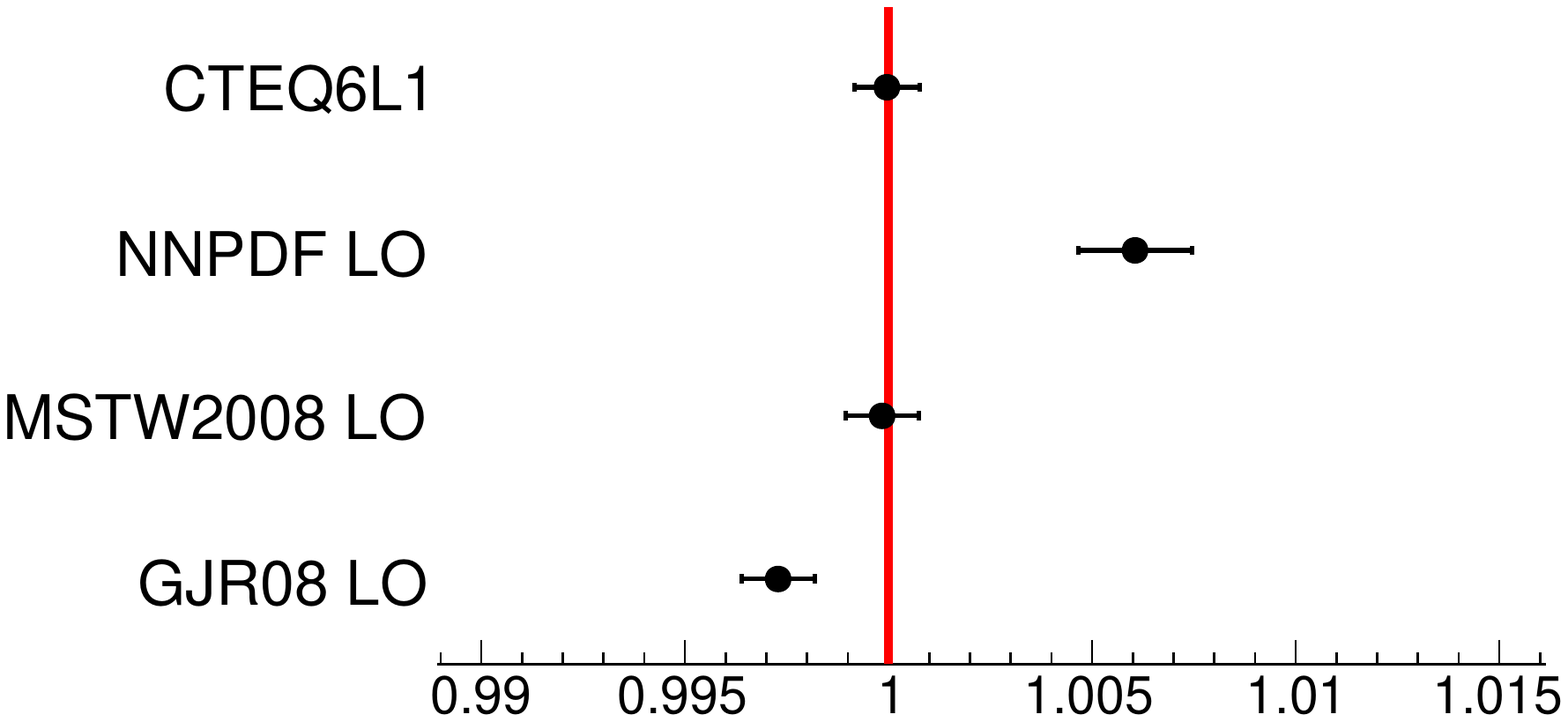}
\caption{Comparison of different pdf set LO asymmetry predictions taking CTEQ6L1 as reference.}\label{fig:pdfchoice}
\end{figure}

A theoretical prediction of $A_{FB}^{b,LHC}$ (and, in particular, of the numerical value of the kinematical constant $k$) has to take into account several experimental issues, thus needing a realistic simulation of the detector features, and in particular of its geometrical properties and of intrinsic cuts applied to the event reconstruction. In such a contest, kinematic cuts on transverse momentum and pseudorapidity of the decaying particles introduce some subtleties in the derivation
  of a direct connection of $A_{FB}^{b, LHC}$ to the LEP asymmetry $A_{FB}^{b}$. 
 \begin{figure} 
\centering{
\includegraphics[scale=.5]{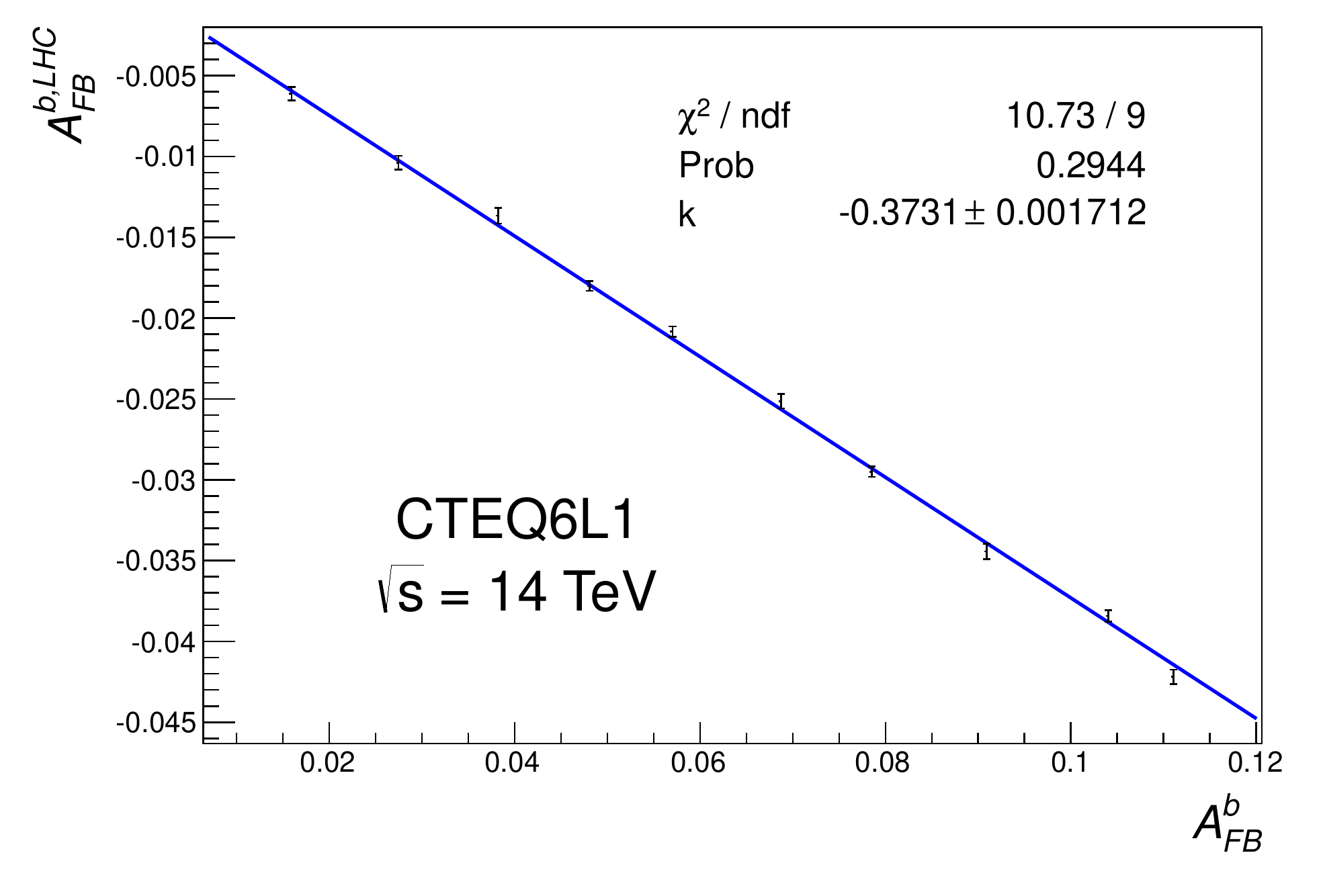}}
\caption{Event level (i.e. without parton showering) dependence of the asymmetry defined in the text on $A_{FB}^{b}$, in a
ficticiously wide range of $A_{FB}^{b}$ values, aiming to prove the \emph{exact} direct proportionality also in the presence of typical kinematic cuts \cite{bib:atlasxsec} on decay products pseudorapidities and transverse momenta. The uncertainty on $k$ is only numerical (see below). }\label{fig:propor}
\end{figure} 
To prove the validity of  \eqref{eq:AFBLHC} also in the presence of a realistic event selection, one can vary fictitiously  $g^{L,R}_{Zb}$ in a wide range of values, determining the corresponding values of $A_{FB}^{b,LHC}$ with usual kinematic cuts. Figure \ref{fig:propor} shows the  results of a simulation with 10 different choices of $g^{L,R}_{Zb}$, including the SM one (for the events simulation we have used CalcHEP \cite{Belyaev:2012qa} and checked good agreement with different event generators).
The particular choice of selection criteria closely follows the one used by ATLAS for the Z-$b$-jets cross section analysis \cite{bib:atlasxsec}. With these assumptions, the kinematical constant $k$  is found to be $-0.37$ at LO. Its QCD scale  dependence has been inspected varying simultaneously the renormalization and factorization scales and computing the corresponding   $A_{FB}^{b,LHC}$ values, Figure \ref{fig:scale}. Similarly the PDF set choice dependence is depicted in Figure \ref{fig:pdfchoice}. The total theoretical uncertainty in both cases is at the $1$ percent level. 
For a detector level simulation one has to choose an appropriate procedure to measure the b-jet charge, that can be achieved adapting the LEP procedure to the LHC case \cite{Akers:1995hn,Krohn:2012fg}.

%\section{The top longitudinal polarization in top-charged Higgs production.}\label{sec:toppol}
\bigskip
{\bf 3.  The top longitudinal polarization in top-charged Higgs production.}
\medskip

\begin{figure}
\centering
\includegraphics[scale=0.6]{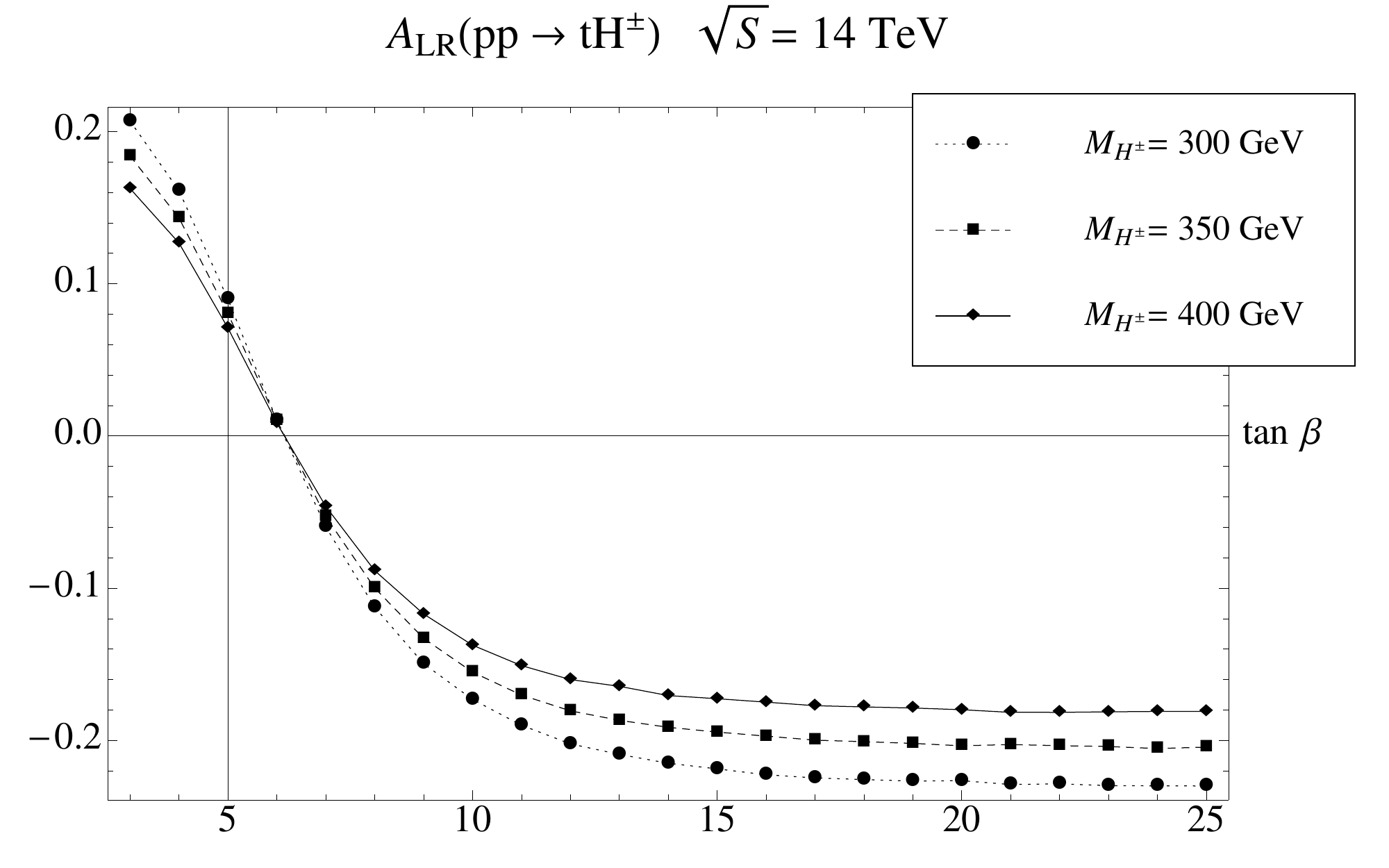}
\caption{Top polarization asymmetry in $tH^{\pm}$ associated production as a function of $\tan \beta$ with tree different assumptions on the charged Higgs mass.}\label{fig:alrtest}
\end{figure}

The previous discussion about $A_{b}$ is not dependent on the assumption of a supersymmetric 
model of New Physics. In particular, there is no impact of SUSY
on $A_b$ if one assumes a heavy enough charged Higgs and sbottoms/stops squarks
which seems to be the case. If Supersymmetry is found, a different asymmetry measurement becomes relevant at LHC, the top  longitudinal polarization asymmetry in top-charged Higgs production. This quantity has been exhaustively discussed in a previous paper~\cite{Baglio:2011ap}, 
where it was shown that its value would essentially mostly
depend on that of the MSSM $\tan \beta$  parameter, and would be almost
 rigorously QCD scale and PDF choice independent. In particular, it was
shown in Ref. \cite{Baglio:2011ap} that varying $\tan \beta$  from approximately one to
approximately ten, the value of the asymmetry changes sign, making an
experimental determination effective even in the presence of a realistic
experimental and theoretical error. For larger $\tan \beta$  values, on the
contrary, the asymmetry remains essentially constant and provides a minor
but still relevant information, and we defer to Ref. \cite{Baglio:2011ap}
for more details.
The relevance of the considered asymmetry appears to us to have been
enormously increased by the latest results on the Higgs boson mass
derived at LHC \cite{ATLAS:2013mma,CMS:yva}. If one wants to retain a MSSM scheme, the
residual range of the Supersymmetric parameters has been greatly reduced.
In particular the allowed values of $\tan \beta$  lie exactly in our ``optimal''
range, roughly from one to ten, with a mass of the charged Higgs in the
$300-600$ GeV range.  Indeed, according to a recent analysis \cite{Djouadi:2013uqa},
while the best fit MSSM point derived from the latest LHC Higgs data
gives $M_{H+} \approx 600$ GeV and $\tan \beta \approx 1$, data are still
in a good agreement with low $\tan \beta$ values and $M_{H+}$ values down to
$300$ GeV (the reason being that the $\chi^2$ is relatively  flat). 
\begin{comment}
(ABDELHAK , MORE COMMENTS). We
have followed in our analysis the indications of a recent paper of Maiani
(Ref \cite{Maiani:2013nga}). In particular, we have assumed the discovery of a charged Higgs
with a mass in the 300-400 GeV range. 
\end{comment}
The variation of the top
polarization asymmetry with $\tan \beta$  in scenarios of this kind is shown in Fig. \ref{fig:alrtest}. 
In our
calculation, we have used the previous results of Ref.\cite{Beccaria:2009my} and have
remained essentially limited to an effective Born approximation. 
The Figure shows the top polarization asymmetry for three different
choices of the Higgs mass: the center of mass energy is $\sqrt{s} = 7~ \text{TeV}$
and, following \cite{Beccaria:2009my,bib:aaa}, the factorization scale $\mu_F$ is set to
$1/6(M_{H^{\pm}}+ m_t)$ to minimize the QCD corrections. The value of the
bottom mass in the Yukawa coupling $t b H^{\pm}$ is evaluated in the
$\overline{MS}$-scheme at the factorization scale.

The main conclusion of our analysis is that a determination of $\tan \beta$  in
the residual range would not request an ``extremely'' precise experimental
measurement. This is a consequence of the fact that a jump from a positive
value of approximately twenty percent to the same value of opposite sign
would not escape a ``reasonable'' determination.

%\section{$A_b$ indications if Supersymmetry is not found at LHC.}\label{sec:Abind}
\bigskip
{\bf 4. $A_b$ indications if Supersymmetry is not found at LHC.}
\medskip

Coming back to the bottom Z process, assuming that Supersymmetry is
found, the proposed determination of $A_b$ from Z polarization becomes
now extremely relevant, given the fact that Supersymmetry would be
unable to explain a discrepancy with the available Standard Model result.
But this asymmetry could also play a fundamental role in the case of a
negative Supersymmetric search at LHC. In particular we shall consider
two opposite cases:

\begin{enumerate}[A)]
\item The $A_b$ value is in disagreement with the Standard Model prediction. 
This result would completely eliminate Supersymmetry, even at a more
powerful proton-proton CERN collider, but would necessarily indicate the
presence of New Physics of non Supersymmetric nature, like that
discussed in some recent papers (see e.g. \cite{Djouadi:2006rk} and references therein).
\item 
The $A_b$ value is in agreement with the Standard Model prediction. This
would leave an ``open door'' for very heavy Supersymmetry, to be searched
at a future more powerful CERN collider, or also exclude effects at LHC
due to a large class of considered New Physics models \cite{Djouadi:2006rk}.
%\textcolor{red}{Assuming a supersymmetric scenario, it is clear that the top quark longitudinal polarization, discussed
%in Section 3, 
%would become extremely interesting to determine $\tan\beta$ and narrow the allowed parameter space of the model.}
\end{enumerate}

The conclusion that we personally think can be derived from our paper is
that, in full generality, a measurement of the Z polarization and top longitudinal asymmetries,
which could be performed at LHC under reasonably expected
experimental conditions, is, to use a mild definition, ``worth''. We are ready
and willing to collaborate with possibly interested experimental teams to
make this project fulfilled.

\bigskip\noindent
{\bf Acknowledgements:} We are very grateful to  Abdelhak Djouadi for  useful discussions and  comments on the manuscript.

\pagebreak


\begin{thebibliography}{00}%{999} 
\begin{comment}
\bibitem{EWFit} ALEPH Collaboration, CDF Collaboration, D0
  Collaboration, DELPHI Collaboration, L3 Collaboration, OPAL
  Collaboration, SLD Collaboration, LEP Electroweak Working Group,
  Tevatron Electroweak Working Group, SLD electroweak heavy flavour
  groups, arXiv:1012.2367v2 [hep-ex] (2011).

%\cite{ATLAS:2012ae}
\bibitem{ATLASHiggs}
  G.~Aad {\it et al.}  [ATLAS Collaboration],
  %``Combined search for the Standard Model Higgs boson using up to 4.9 fb-1 of pp collision data at sqrt(s) = 7 TeV with the ATLAS detector at the LHC,''
  Phys.\ Lett.\ B {\bf 710} (2012) 49
  [arXiv:1202.1408 [hep-ex]].
  %%CITATION = ARXIV:1202.1408;%%

%\cite{Chatrchyan:2012tx}
\bibitem{CMSHiggs}
  S.~Chatrchyan {\it et al.}  [CMS Collaboration],
  %``Combined results of searches for the standard model Higgs boson in pp collisions at sqrt(s) = 7 TeV,''
  arXiv:1202.1488 [hep-ex].
  %%CITATION = ARXIV:1202.1488;%%

%\cite{ALEPH:2005ab}
\bibitem{LEPEWWG}
  [ALEPH and DELPHI and L3 and OPAL and SLD and LEP Electroweak Working Group and SLD Electroweak Group and SLD Heavy Flavour Group Collaborations],
  %``Precision electroweak measurements on the $Z$ resonance,''
  Phys.\ Rept.\  {\bf 427} (2006) 257
  [hep-ex/0509008].
  %%CITATION = HEP-EX/0509008;%%

%\cite{Chanowitz:2001bv}
\bibitem{ZbbBSM1}
  M.~S.~Chanowitz,
  %``The Z ---> anti-b b decay asymmetry: Lose-lose for the standard model,''
  Phys.\ Rev.\ Lett.\  {\bf 87} (2001) 231802
  [hep-ph/0104024].
  %%CITATION = HEP-PH/0104024;%%

%\cite{He:2003qv}
\bibitem{ZbbBSM2}
  X.~-G.~He and G.~Valencia,
  %``A**b(FB) and R(b) at LEP and new right-handed gauge bosons,''
  Phys.\ Rev.\ D {\bf 68} (2003) 033011
  [hep-ph/0304215].
  %%CITATION = HEP-PH/0304215;%%

%\cite{DaRold:2010as}
\bibitem{ZbbBSM3}
  L.~Da Rold,
  %``Solving the $A_{FB}^b$ anomaly in natural composite models,''
  JHEP {\bf 1102} (2011) 034
  [arXiv:1009.2392 [hep-ph]].
  %%CITATION = ARXIV:1009.2392;%%

%\cite{Cao:2008rc}
\bibitem{ZbbMSSM}
  J.~Cao and J.~M.~Yang,
  %``Anomaly of Zb anti-b coupling revisited in MSSM and NMSSM,''
  JHEP {\bf 0812} (2008) 006
  [arXiv:0810.0751 [hep-ph]].
  %%CITATION = ARXIV:0810.0751;%%

%\cite{Campbell:2003dd}
\bibitem{ZbNloQCD}
  J.~M.~Campbell, R.~K.~Ellis, F.~Maltoni and S.~Willenbrock,
  %``Associated production of a $Z$ Boson and a single heavy quark jet,''
  Phys.\ Rev.\ D {\bf 69} (2004) 074021
  [hep-ph/0312024].
  %%CITATION = HEP-PH/0312024;%%



%\cite{Pumplin:2002vw}
\bibitem{CTEQ}
  J.~Pumplin, D.~R.~Stump, J.~Huston, H.~L.~Lai, P.~M.~Nadolsky and W.~K.~Tung,
  %``New generation of parton distributions with uncertainties from global QCD analysis,''
  JHEP {\bf 0207} (2002) 012
  [hep-ph/0201195].
  %%CITATION = HEP-PH/0201195;%%

\bibitem{AsymTheory} Placeholder


%\cite{Denner:2011vu}
\bibitem{Denner}
  A.~Denner, S.~Dittmaier, T.~Kasprzik and A.~Muck,
  %``Electroweak corrections to dilepton + jet production at hadron colliders,''
  JHEP {\bf 1106} (2011) 069
  [arXiv:1103.0914 [hep-ph]].
  %%CITATION = ARXIV:1103.0914;%%

%\cite{Martin:2009iq}
\bibitem{MSTW}
  A.~D.~Martin, W.~J.~Stirling, R.~S.~Thorne and G.~Watt,
  %``Parton distributions for the LHC,''
  Eur.\ Phys.\ J.\ C {\bf 63} (2009) 189
  [arXiv:0901.0002 [hep-ph]].
  %%CITATION = ARXIV:0901.0002;%%

%\cite{Ball:2011mu}
\bibitem{NNPDF}
  R.~D.~Ball, V.~Bertone, F.~Cerutti, L.~Del Debbio, S.~Forte, A.~Guffanti, J.~I.~Latorre and J.~Rojo {\it et al.},
  %``Impact of Heavy Quark Masses on Parton Distributions and LHC Phenomenology,''
  Nucl.\ Phys.\ B {\bf 849} (2011) 296
  [arXiv:1101.1300 [hep-ph]].
  %%CITATION = ARXIV:1101.1300;%%


%\cite{Aad:2011jn}
\bibitem{ATLASZb}
  G.~Aad {\it et al.}  [ATLAS Collaboration],
  %``Measurement of the cross-section for b-jets produced in association with a Z boson at sqrt(s)=7 TeV with the ATLAS detector,''
  Phys.\ Lett.\ B {\bf 706} (2012) 295
  [arXiv:1109.1403 [hep-ex]].
  %%CITATION = ARXIV:1109.1403;%%


%\cite{Beccaria:2008av}
\bibitem{tW1}
  M.~Beccaria, C.~M.~Carloni Calame, G.~Macorini, E.~Mirabella, F.~Piccinini, F.~M.~Renard and C.~Verzegnassi,
  %``A Complete one-loop calculation of electroweak supersymmetric effects in t-channel single top production at CERN LHC,''
  Phys.\ Rev.\ D {\bf 77} (2008) 113018
  [arXiv:0802.1994 [hep-ph]].
  %%CITATION = ARXIV:0802.1994;%%
  
%\cite{Beccaria:2007tc}
\bibitem{tW2}
  M.~Beccaria, C.~M.~Carloni Calame, G.~Macorini, G.~Montagna, F.~Piccinini, F.~M.~Renard and C.~Verzegnassi,
  %``A Complete one-loop description of associated tW production at LHC and a search for possible genuine supersymmetric effects,''
  Eur.\ Phys.\ J.\ C {\bf 53} (2008) 257
  [arXiv:0705.3101 [hep-ph]].
  %%CITATION = ARXIV:0705.3101;%%
  
%\cite{Denner:2009gj}
\bibitem{Wjets}
  A.~Denner, S.~Dittmaier, T.~Kasprzik and A.~Muck,
  %``Electroweak corrections to W + jet hadroproduction including leptonic W-boson decays,''
  JHEP {\bf 0908} (2009) 075
  [arXiv:0906.1656 [hep-ph]].
  %%CITATION = ARXIV:0906.1656;%%

%\cite{Beccaria:2003sq}
\bibitem{HighELog1}
  M.~Beccaria, F.~M.~Renard, S.~Trimarchi and C.~Verzegnassi,
  %``Higgs production in the 1-TeV domain: Logarithmic Sudakov expansion and tan theta_{\bar{l}} measurement analysis,''
  LC-TH-2003-070.
  %%CITATION = LC-TH-2003-070;%%

%\cite{Beccaria:2003yn}
\bibitem{HighELog2}
  M.~Beccaria, M.~Melles, F.~M.~Renard, S.~Trimarchi and C.~Verzegnassi,
  %``Sudakov expansions at one loop and beyond for charged scalar and fermion pair production in SUSY models at future linear colliders,''
  Int.\ J.\ Mod.\ Phys.\ A {\bf 18} (2003) 5069
  [hep-ph/0304110].
  %%CITATION = HEP-PH/0304110;%%

%\cite{Gounaris:2011pp}
\bibitem{AugSud}
  G.~J.~Gounaris and F.~M.~Renard,
  %``Supersimplicity: a remarkable high energy SUSY property,''
  Acta Phys.\ Polon.\ B {\bf 42} (2011) 2107
  [arXiv:1106.2707 [hep-ph]].
  %%CITATION = ARXIV:1106.2707;%%

%\cite{Baglio:2011ap}
\bibitem{tH-}
  J.~Baglio, M.~Beccaria, A.~Djouadi, G.~Macorini, E.~Mirabella, N.~Orlando, F.~M.~Renard and C.~Verzegnassi,
  %``The Left-Right asymmetry of top quarks in associated top-charged Higgs bosons at the LHC as a probe of the tan$\theta_{\bar{l}}$ parameter,''
  Phys.\ Lett.\ B {\bf 705} (2011) 212
  [arXiv:1109.2420 [hep-ph]].
  %%CITATION = ARXIV:1109.2420;%%
  \end{comment}
%1

 %\cite{Blondel:1987gp}
\bibitem{Blondel:1987gp}
  A.~Blondel, B.~W.~Lynn, F.~M.~Renard and C.~Verzegnassi,
  %``Precision Measurements Of Final State Weak Coupling From Polarized Electron - Positron Annihilation,''
  Nucl.\ Phys.\ B {\bf 304} (1988) 438.
  %%CITATION = NUPHA,B304,438;%%
  %55 citations counted in INSPIRE as of 18 Mar 2013
  
  %\cite{Lynn:1986ir}
\bibitem{Lynn:1986ir}
  B.~W.~Lynn and C.~Verzegnassi,
  %``LONGITUDINAL e- BEAM POLARIZATION ASYMMETRY IN e+ e- ---> HADRONS,''
  Phys.\ Rev.\ D {\bf 35} (1987) 3326.
  %%CITATION = PHRVA,D35,3326;%%
  %53 citations counted in INSPIRE as of 18 Mar 2013
  
  
  
  %\cite{Abe:1994tv}
\bibitem{Abe:1994tv}
  K.~Abe {\it et al.}  [SLD Collaboration],
  %``Measuring A(b) with polarized beams at SLC,''
 % In *Meribel les Allues 1994, Proceedings, '94 electroweak interactions and unified theories* 123-129, and SLAC Stanford - 
 SLAC-PUB-6513\ (94/05,rec.Aug.) 9 p
  %2 citations counted in INSPIRE as of 19 Mar 2013

%\cite{Abe:2000dq}
\bibitem{Abe:2000dq}
  K.~Abe {\it et al.}  [SLD Collaboration],
  %``A High precision measurement of the left-right Z boson cross-section asymmetry,''
  Phys.\ Rev.\ Lett.\  {\bf 84} (2000) 5945
  [hep-ex/0004026].
  %%CITATION = HEP-EX/0004026;%%
  %75 citations counted in INSPIRE as of 12 Jul 2013

%\cite{Baak:2012kk}
\bibitem{Baak:2012kk}
  M.~Baak, M.~Goebel, J.~Haller, A.~Hoecker, D.~Kennedy, R.~Kogler, K.~Moenig and M.~Schott {\it et al.},
  %``The Electroweak Fit of the Standard Model after the Discovery of a New Boson at the LHC,''
  Eur.\ Phys.\ J.\ C {\bf 72} (2012) 2205
  [arXiv:1209.2716 [hep-ph]].
  %%CITATION = ARXIV:1209.2716;%%
  %29 citations counted in INSPIRE as of 19 Mar 2013

%\cite{ALEPH:2005ab}
\bibitem{ALEPH:2005ab}
  S.~Schael {\it et al.}  [ALEPH and DELPHI and L3 and OPAL and SLD and LEP Electroweak Working Group and SLD Electroweak Group and SLD Heavy Flavour Group Collaborations],
  %``Precision electroweak measurements on the $Z$ resonance,''
  Phys.\ Rept.\  {\bf 427} (2006) 257
  [hep-ex/0509008].
  %%CITATION = HEP-EX/0509008;%%
  %826 citations counted in INSPIRE as of 07 Aug 2013

\bibitem{ref:MSSMunable}%\cite{Cao:2008rc}
  J.~Cao and J.~M.~Yang,
  %``Anomaly of Zb anti-b coupling revisited in MSSM and NMSSM,''
  JHEP {\bf 0812} (2008) 006
  [arXiv:0810.0751 [hep-ph]].
  %%CITATION = ARXIV:0810.0751;%%
  %30 citations counted in INSPIRE as of 12 Jul 2013

%\cite{Freitas:2012sy}
\bibitem{Freitas:2012sy}
  A.~Freitas and Y.~-C.~Huang,
  %``Electroweak two-loop corrections to sin^2{\theta}(eff,bb) and R(b) using numerical Mellin-Barnes integrals,''
  JHEP {\bf 1208} (2012) 050
  [arXiv:1205.0299 [hep-ph]].
  %%CITATION = ARXIV:1205.0299;%%
  %12 citations counted in INSPIRE as of 19 Mar 2013

  
  %\cite{Mirkes:1992hu}
%\bibitem{Zpol1}
%  E.~Mirkes,
  %``Angular decay distribution of leptons from W bosons at NLO in hadronic collisions,''
%  Nucl.\ Phys.\ B {\bf 387} (1992) 3.
  %%CITATION = NUPHA,B387,3;%%

%2
%\cite{Mirkes:1994dp}
%\bibitem{Zpol2}
%  E.~Mirkes and J.~Ohnemus,
  %``Angular distributions of Drell-Yan lepton pairs at the Tevatron: Order $\alpha-s^{2}$ corrections and Monte Carlo studies,''
%  Phys.\ Rev.\ D {\bf 51} (1995) 4891
%  [hep-ph/9412289].
  %%CITATION = HEP-PH/9412289;%%

%3  
%\bibitem{Zpol3}
  %\cite{Mirkes:1994eb}
%\bibitem{Mirkes:1994eb}
%  E.~Mirkes and J.~Ohnemus,
  %``$W$ and $Z$ polarization effects in hadronic collisions,''
%  Phys.\ Rev.\ D {\bf 50} (1994) 5692
%  [hep-ph/9406381].
  %%CITATION = HEP-PH/9406381;%%
  %28 citations counted in INSPIRE as of 18 Mar 2013

%4  
%\cite{Jacob:1959at}
%\bibitem{Jacob:1959at}
%  M.~Jacob and G.~C.~Wick,
  %``On the general theory of collisions for particles with spin,''
%  Annals Phys.\  {\bf 7} (1959) 404
%   [Annals Phys.\  {\bf 281} (2000) 774].
  %%CITATION = APNYA,7,404;%%

%5

%\cite{Wick:1962zz}
%\bibitem{Wick:1962zz}
%  G.~C.~Wick,
  %``Angular momentum states for three relativistic particles,''
%  Annals Phys.\  {\bf 18} (1962) 65.
  %%CITATION = APNYA,18,65;%%

%6
%\cite{Choi:1994nv}
%\bibitem{Choi:1994nv}
%  S.~Y.~Choi,
  %``Probing the weak boson sector in gamma e ---> Z e,''
%  Z.\ Phys.\ C {\bf 68} (1995) 163
%  [hep-ph/9412300].
  %%CITATION = HEP-PH/9412300;%%
  
  %7
  
 %\cite{Stirling:2013muo}
\bibitem{Stirling:2013muo}
  J.~Stirling and E.~Vryonidou,
  %``Polarisation of electroweak gauge bosons at the LHC,''
  arXiv:1302.1365 [hep-ph].
  %%CITATION = ARXIV:1302.1365;%%
  
  %8
  
 %\cite{ATLAS:2012au}
%\bibitem{ATLAS:2012au}
%  G.~Aad {\it et al.}  [ATLAS Collaboration],
  %``Measurement of the polarisation of $W$ bosons produced with large transverse momentum in $pp$ collisions at $\sqrt{s}=7$ TeV with the ATLAS experiment,''
%  Eur.\ Phys.\ J.\ C {\bf 72} (2012) 2001
%  [arXiv:1203.2165 [hep-ex]].
  %%CITATION = ARXIV:1203.2165;%%
  %12 citations counted in INSPIRE as of 18 Mar 2013

%9

%\cite{Chatrchyan:2011ig}
%\bibitem{Chatrchyan:2011ig}
%  S.~Chatrchyan {\it et al.}  [CMS Collaboration],
  %``Measurement of the Polarization of W Bosons with Large Transverse Momenta in W+Jets Events at the LHC,''
%  Phys.\ Rev.\ Lett.\  {\bf 107} (2011) 021802
%  [arXiv:1104.3829 [hep-ex]].
  %%CITATION = ARXIV:1104.3829;%%
  %28 citations counted in INSPIRE as of 18 Mar 2013

%10  
  
  %\cite{Beccaria:2012xw}
\bibitem{Beccaria:2012xw}
  M.~Beccaria, N.~Orlando, G.~Panizzo, F.~M.~Renard and C.~Verzegnassi,
  %``The Relevance of polarized bZ production at LHC,''
  Phys.\ Lett.\ B {\bf 713} (2012) 457
  [arXiv:1204.5315 [hep-ph]].
  %%CITATION = ARXIV:1204.5315;%%
  %1 citations counted in INSPIRE as of 18 Mar 2013

%11  


%\cite{Belyaev:2012qa}
\bibitem{Belyaev:2012qa}
  A.~Belyaev, N.~D.~Christensen and A.~Pukhov,
  %``CalcHEP 3.4 for collider physics within and beyond the Standard Model,''
  Comput.\ Phys.\ Commun.\  {\bf 184} (2013) 1729
  [arXiv:1207.6082 [hep-ph]].
  %%CITATION = ARXIV:1207.6082;%%
  %47 citations counted in INSPIRE as of 18 Jul 2013


%\cite{Aad:2011jn}
\bibitem{bib:atlasxsec}
  G.~Aad {\it et al.}  [ATLAS Collaboration],
  %``Measurement of the cross-section for $b^-$ jets produced in association with a $Z$ boson at $\sqrt{s}=7$ TeV with the ATLAS detector,''
  Phys.\ Lett.\ B {\bf 706} (2012) 295
  [arXiv:1109.1403 [hep-ex]].
  %%CITATION = ARXIV:1109.1403;%%
  %31 citations counted in INSPIRE as of 08 Apr 2013

%\cite{Akers:1995hn}
\bibitem{Akers:1995hn}
  R.~Akers {\it et al.}  [OPAL Collaboration],
  %``A Measurement of the forward - backward asymmetry of e+ e- --> b anti-b by applying a jet charge algorithm to lifetime tagged events,''
  Z.\ Phys.\ C {\bf 67} (1995) 365.
  %%CITATION = ZEPYA,C67,365;%%
  %39 citations counted in INSPIRE as of 06 Aug 2013
%\cite{Krohn:2012fg}
\bibitem{Krohn:2012fg}
  D.~Krohn, M.~D.~Schwartz, T.~Lin and W.~J.~Waalewijn,
  %``Jet Charge at the LHC,''
  Phys.\  Rev.\  Lett.\  {\bf 110} (2013) 212001
  [arXiv:1209.2421 [hep-ph]].
  %%CITATION = ARXIV:1209.2421;%%
  %8 citations counted in INSPIRE as of 06 Aug 2013

%\cite{Baglio:2011ap}
\bibitem{Baglio:2011ap}
  J.~Baglio, M.~Beccaria, A.~Djouadi, G.~Macorini, E.~Mirabella, N.~Orlando, F.~M.~Renard and C.~Verzegnassi,
  %``The Left-Right asymmetry of top quarks in associated top-charged Higgs bosons at the LHC as a probe of the tan$\beta$ parameter,''
  Phys.\ Lett.\ B {\bf 705} (2011) 212
  [arXiv:1109.2420 [hep-ph]].
  %%CITATION = ARXIV:1109.2420;%%
  %10 citations counted in INSPIRE as of 18 Jul 2013  

%\cite{ATLAS:2013mma}
\bibitem{ATLAS:2013mma}
  [ATLAS Collaboration],
  %``Combined measurements of the mass and signal strength of the Higgs-like boson with the ATLAS detector using up to 25 fb$^{-1}$ of proton-proton collision data,''
  ATLAS-CONF-2013-014.
  %%CITATION = ATLAS-CONF-2013-014;%%
  %37 citations counted in INSPIRE as of 18 Jul 2013

%\cite{CMS:yva}
\bibitem{CMS:yva}
  [CMS Collaboration],
  %``Combination of standard model Higgs boson searches and measurements of the properties of the new boson with a mass near 125 GeV,''
  CMS-PAS-HIG-13-005.
  %%CITATION = CMS-PAS-HIG-13-005;%%
  %47 citations counted in INSPIRE as of 18 Jul 2013


%\cite{Maiani:2013nga}
%\bibitem{Maiani:2013nga}
%  L.~Maiani, A.~Polosa and V.~Riquer,
  %``Bounds to the Higgs Sector Masses in Minimal Supersymmetry from LHC Data,''
%  Phys.\ Lett.\ B {\bf 724} (2013) 274
%  [arXiv:1305.2172 [hep-ph]].
  %%CITATION = ARXIV:1305.2172;%%
  %1 citations counted in INSPIRE as of 18 Jul 2013

%\cite{Djouadi:2013uqa}
\bibitem{Djouadi:2013uqa}
  A.~Djouadi, L.~Maiani, G.~Moreau, A.~Polosa, J.~Quevillon and V.~Riquer,
  %``The post-Higgs MSSM scenario: Habemus MSSM?,''
  arXiv:1307.5205 [hep-ph].
  %%CITATION = ARXIV:1307.5205;%%
  %1 citations counted in INSPIRE as of 06 Aug 2013

%\cite{Beccaria:2009my}
\bibitem{Beccaria:2009my}
  M.~Beccaria, G.~Macorini, L.~Panizzi, F.~M.~Renard and C.~Verzegnassi,
  %``Associated production of charged Higgs and top at LHC: The Role of the complete electroweak supersymmetric contribution,''
  Phys.\ Rev.\ D {\bf 80} (2009) 053011
  [arXiv:0908.1332 [hep-ph]].
  %%CITATION = ARXIV:0908.1332;%%
  %15 citations counted in INSPIRE as of 18 Jul 2013  
  
  
\bibitem{bib:aaa} 
  T.~Plehn, 
  Phys.\ Rev.\ D {\bf 67} (2003) 014018; 
  S.\ Zhou, 
  Phys.\ Rev.\ D {\bf 67} (2003) 075006.

%\cite{Djouadi:2006rk}
\bibitem{Djouadi:2006rk}
  A.~Djouadi, G.~Moreau and F.~Richard,
  %``Resolving the A(FB)**b puzzle in an extra dimensional model with an extended gauge structure,''
  Nucl.\ Phys.\ B {\bf 773} (2007) 43
  [hep-ph/0610173].
  %%CITATION = HEP-PH/0610173;%%
  %74 citations counted in INSPIRE as of 07 Aug 2013
  
  
\end{thebibliography}
\end{document}